\documentstyle[12pt]{article}

\input{tcilatex}

\begin{document}

\title{2D H - atom in an arbitrary magnetic field via pseudoperturbation expansions
through the quantum number $l$ }
\author{Omar Mustafa$^{\dagger}$ and Maen Odeh \\
Department of Physics, Eastern Mediterranean University\\
G. Magusa, North Cyprus, Mersin 10 - Turkey\\
${\dagger}$ email: omustafa@mozart.emu.edu.tr\\
}
\maketitle

\begin{abstract}
{\small The pseudoperturbative shifted - $l$ expansion technique (PSLET) is
introduced to determine nodeless states of the 2D Schr\"odinger equation
with arbitrary cylindrically symmetric potentials. Exact energy eigenvalues
and eigenfunctions for the 2D Coulomb and harmonic oscillator potentials are
reproduce. Moreover, exact energy eigenvalues, compared to those obtained by
numerical solution [11], were obtained for the hybrid of the 2D Coulomb and
oscillator potentials.}
\end{abstract}

\newpage\ \ 

\section{\ \ \ Introduction}

Recent advances in nanofabrication technology have made possible to create
low dimensional structures [ 1-4 and references therein] such as quantum
wells, wires and dots. In almost most of problems concerning such structures
the eigenspectrum of the 2D- Schr\H{o}dinger equation is investigated. For
example, the 2D hydrogenic energy levels in a constant magnetic field of
arbitrary strength has been a subject of numerous theoretical and
experimental investigations [1-14].

Taut [6] has found analytical solutions for this problem. However, no
solutions for nodeless states were found. Martin et al [10] have used a two
- point quasifractional approximation and found excellent interpolation
between the weak and strong magnetic field perturbation expansions.
MacDonald and Ritchie [8] have however used a two - point Pad\'{e}
approximation. The curves obtained from different Pad\'{e} approximates are
very different. No regular pattern appears and the results become
unreliable. More detailed discussion can be found in ref [10]. Zhu et al [9]
have used a power series expansion method. Their results compared
excellently with the direct numerical integration [12] and perturbation [8]
methods in the weak field regime. However, neither Zhu's nor direct
numerical integration results seem to approach the perturbation results at
the strong field limit. Villalba and Pino [11] have used a finite -
difference scheme with a linear mesh of up to 2000 points and failed to
provide a good estimation of the ground state for the 2D hydrogen atom. On
the other hand, their variational solution obtained using hydrogen basis is
in good agreement with that of Martin et al [10] at the weak field limit.
Using the oscillator basis, their results were in good agreement at the
strong field limit. Their solutions could not therefore provide information
about how the energy shifts in the intermediate range of the magnetic field.
Mustafa [5] and Quiroga et al [7] have used the shifted N - expansion
technique (SLNT) and their results were in good agreements with those of
Martin et al [10].

In this work, section two considers the 2D Shcr\H{o}dinger equation with an
arbitrary cylindrically symmetric potential. By introducing a technique,
which we will call Pseudoperturbative Shifted - $l$ Expansion Technique
(PSLET), \ we obtain analytical expressions for both eigenenergies and
eigenfunctions for the 2D Schr\H{o}dinger equation. In section 3, we test
our method for both 2D Coulomb and harmonic oscillator potentials, and the
results found to be exact. We also consider the 2D hydrogenic energy levels
in a constant magnetic field of arbitrary strength. PSLET results are shown
to compare excellently with the perturbation [8], direct numerical
integration [12], and series expansion [9] methods. They are also shown to
be in exact agreements with the numerical solution of the 2D Schr\"{o}dinger
equation [10]. We conclude with remarkes in section 4.

\section{ Theory of 2D-PSLET}

The Schr\"odinger equation for an arbitrary cylindrically symmetric
potential $V(\rho)$ ( in $\hbar = 2m = 1$ units) is\newline
\begin{equation}
\left[-\frac{d^{2}}{d\rho^{2}}+\frac{4l^2-1}{4\rho^{2}}+V(\rho)\right]%
\Psi(\rho)=E\Psi(\rho),
\end{equation}
\newline
where $\rho^2 = x^2 + y^2$, $l=|m|$ and $m$ is the magnetic quantum number.
If $l$ is shifted through the relation $\bar{l} = l - \beta$, Eq.(1) becomes%
\newline
\begin{equation}
\left\{-\frac{d^{2}}{d\rho^{2}}+\tilde{V}(\rho)\right\}\Psi(\rho)=E\Psi(%
\rho),
\end{equation}
\newline
with\newline
\begin{equation}
\tilde{V}(\rho)=\frac{(\bar{l}+\beta+1/2)(\bar{l}+\beta-1/2)}{\rho^{2}} +%
\frac{\bar{l}^2}{Q}V(\rho).
\end{equation}
\newline
Where Q is a scale to be determined below and set equal to $\bar{l}^2$ at
the end of the calculations, and $\beta$ is a suitable shift to be
determined and is introduced partly to avoid the trivial case $l=0$.

The systematic procedure of PSLET starts with making use of Taylor's theorem
and expanding (2) about an arbitrary point ( for now) on the $\rho $ - axis.
It is convenient then to transform the coordinates in (2) via the relation%
\newline
\begin{equation}
x=\bar{l}^{1/2}(\rho -\rho _{o})/\rho _{o},
\end{equation}
\newline
where $\rho _{o}$ is our arbitrary point. Expansions about $x=0$ yield%
\newline
\begin{equation}
\left[ -\frac{d^{2}}{dx^{2}}+\tilde{V}(x(\rho ))\right] \Psi (x)=\frac{\rho
_{o}^{2}}{\bar{l}}E\Psi (x),
\end{equation}
\newline
where\newline
\begin{eqnarray}
\tilde{V}(x(\rho )) &=&\rho _{o}^{2}\bar{l}\left[ \frac{1}{\rho _{o}^{2}}+%
\frac{V(\rho _{o})}{Q}\right] +\bar{l}^{1/2}\left[ -2x+\frac{V^{^{\prime
}}(\rho _{o})\rho _{o}^{3}x}{Q}\right]  \nonumber \\
&+&\left[ 3x^{2}+\frac{V^{^{\prime \prime }}(\rho _{o})\rho _{o}^{4}x^{2}}{2Q%
}\right] +2\beta \sum_{n=1}^{\infty }(-1)^{n}(n+1)x^{n}\bar{l}^{-n/2} 
\nonumber \\
&+&\sum_{n=3}^{\infty }\left[ (-1)^{n}(n+1)x^{n}+\left( \frac{d^{n}V(\rho
_{o})}{d\rho _{o}^{n}}\right) \frac{\rho _{o}^{2}(\rho _{o}x)^{n}}{n!Q}%
\right] \bar{l}^{-(n-2)/2}  \nonumber \\
&+&(\beta ^{2}-1/4)\sum_{n=0}^{\infty }(-1)^{n}(n+1)x^{n}\bar{l}%
^{-(n+2)/2}+2\beta .
\end{eqnarray}
\newline
It is convenient to expand $E$ as\newline
\begin{equation}
E=\sum_{n=-2}^{\infty }E^{(n)}\bar{l}^{-n}.
\end{equation}
\newline
where the prime of $V(\rho _{o})$ denotes derivative with respect to $\rho
_{o}$. Equation (6) when compared to Schr\"{o}dinger equation for one -
dimensional anharmonic oscillator\newline
\begin{equation}
\left[ -\frac{d^{2}}{dx^{2}}+\frac{1}{4}w^{2}x^{2}+\varepsilon _{o}+P(x)%
\right] X_{n_{\rho }}(x)=\lambda _{n_{\rho }}X_{n_{\rho }}(x),
\end{equation}
\newline
where $P(x)$ is a perturbation term and $\varepsilon _{o}$ is a constant,
implies\newline
\begin{equation}
\varepsilon _{o}=\bar{l}\left[ 1+\frac{\rho _{o}^{2}V(\rho _{o})}{Q}\right]
+2\beta +\frac{(\beta ^{2}-1/4)}{\bar{l}},
\end{equation}
\newline
\begin{equation}
\lambda _{n_{\rho }}=\varepsilon _{o}+(n_{\rho }+1/2)w+\lambda ^{(0)}/\bar{l}%
+\sum_{n=2}^{\infty }\lambda ^{(n-1)}\bar{l}^{-n},
\end{equation}
\newline
and\newline
\begin{equation}
\lambda _{n_{\rho }}=\rho _{o}^{2}\sum_{n=-2}^{\infty }E^{(n)}\bar{l}%
^{-(n+1)}.
\end{equation}
\newline
Equations (10) and (11) yield\newline
\begin{equation}
E^{(-2)}=\frac{1}{\rho _{o}^{2}}+\frac{V(\rho _{o})}{Q}
\end{equation}
\newline
\begin{equation}
E^{(-1)}=\frac{1}{\rho _{o}^{2}}\left[ 2\beta +(n_{\rho }+1/2)w\right]
\end{equation}
\newline
\begin{equation}
E^{(0)}=\frac{1}{\rho _{o}^{2}}\left[ \beta ^{2}-1/4+\lambda ^{(0)}\right]
\end{equation}
\newline
\begin{equation}
E^{(n)}=\lambda ^{(n)}/\rho _{o}^{2}~~;~~~~n\geq 1.
\end{equation}
\newline
Here $\rho _{o}$ is chosen to minimize $E^{(-2)}$, i. e.\newline
\begin{equation}
\frac{dE^{(-2)}}{d\rho _{o}}=0~~~~and~~~~\frac{d^{2}E^{(-2)}}{d\rho _{o}^{2}}%
>0,
\end{equation}
\newline
which in turn gives, with $\bar{l}=\sqrt{Q}$,\newline
\begin{equation}
l-\beta =\sqrt{\frac{\rho _{o}^{3}V^{^{\prime }}(\rho _{o})}{2}}.
\end{equation}
\newline

The shifting parameter $\beta $ is determined by choosing the next leading
correction to the energy eigenvalue, $E^{(-1)}$, to vanish. This choice is
physically motivated by requiring the agreement between 2D-PSLET eigenvalues
and eigenfunctions and the exact analytical eigenvalues and eigenfunctions
for both the Coulomb and harmonic oscillator potentials. Hence\newline
\begin{equation}
\beta =-\frac{1}{2}\left[ (n_{\rho }+\frac{1}{2})w\right] ,
\end{equation}
\newline
where\newline
\begin{equation}
w=2\sqrt{3+\frac{\rho _{o}V^{^{\prime \prime }}(\rho _{o})}{V^{^{\prime
}}(\rho _{o})}}.
\end{equation}
\newline

Then equation (6) becomes\newline
\begin{equation}
\tilde{V}(x(\rho ))=\rho _{o}^{2}\bar{l}E^{(-2)}+\sum_{n=0}^{\infty
}v^{(n)}(x)\bar{l}^{-n/2},
\end{equation}
\newline
where\newline
\begin{equation}
v^{(0)}(x)=\frac{1}{4}w^{2}x^{2}+2\beta ,
\end{equation}
\newline
\begin{equation}
v^{(1)}(x)=-4\beta x-4x^{3}+\frac{\rho _{o}^{5}V^{^{\prime \prime \prime
}}(\rho _{o})}{6Q}x^{3},
\end{equation}
\newline
\begin{equation}
v^{(2)}(x)=(\beta ^{2}-1/4)+6\beta x^{2}+\left( 5+\frac{\rho
_{o}^{6}V^{^{\prime \prime \prime \prime }}(\rho _{o})}{24Q}\right) x^{4},
\end{equation}
\newline
and for $n\geq 3$\newline
\begin{eqnarray}
v^{(n)}(x) &=&(-1)^{n}2\beta (n+1)x^{n}+(-1)^{n}(\beta ^{2}-1/4)(n-1)x^{n-2}
\nonumber \\
&+&\left[ (-1)^{n}(n+3)+\frac{\rho _{o}^{n+4}}{Q(n+2)!}\frac{d^{n+2}V(\rho
_{o})}{d\rho _{o}^{n+2}}\right] x^{n+2}.
\end{eqnarray}
\newline
Equation (5) thus becomes\newline
\begin{eqnarray}
&&\left[ -\frac{d^{2}}{dx^{2}}+\sum_{n=0}^{\infty }v^{(n)}\bar{l}^{-n/2}%
\right] \Psi _{n_{\rho }}(x)=  \nonumber \\
&&\left[ \frac{1}{\bar{l}}\left( \beta ^{2}-\frac{1}{4}+\lambda
^{(0)}\right) +\sum_{n=2}^{\infty }\lambda ^{(n-1)}\bar{l}^{-n}\right] \Psi
_{n_{\rho }}(x).
\end{eqnarray}
\newline

For nodeless unnormalized wave functions $n_{\rho }=0$ and \newline
\begin{equation}
\Psi _{0}(x(\rho ))=exp(U(x)),
\end{equation}
\newline
which when substituted in equation (25) yields\newline
\begin{eqnarray}
-[U^{^{\prime \prime }}(x)+U^{^{\prime }}(x)U^{^{\prime
}}(x)]+\sum_{n=0}^{\infty }v^{(n)}(x)\bar{l}^{-n/2} &=&\frac{1}{\bar{l}}%
\left( \beta ^{2}-\frac{1}{4}+\lambda ^{(0)}\right)  \nonumber \\
&&+\sum_{n=2}^{\infty }\lambda ^{(n-1)}\bar{l}^{-n},
\end{eqnarray}
\newline
where prime of $U(x)$ denotes derivative with respect to $x$. It is evident
that this equation admits solution of the form \newline
\begin{equation}
U^{^{\prime }}(x)=\sum_{n=0}^{\infty }U^{(n)}(x)\bar{l}^{-n/2}+\sum_{n=0}^{%
\infty }G^{(n)}(x)\bar{l}^{-(n+1)/2},
\end{equation}
\newline
where\newline
\begin{equation}
U^{(n)}(x)=\sum_{j=0}^{n+1}D_{j,n}x^{2j-1}~~~~;~~~D_{0,n}=0,
\end{equation}
\newline
\begin{equation}
G^{(n)}(x)=\sum_{j=0}^{n+1}C_{j,n}x^{2j}.
\end{equation}
\newline
Substituting equations (28) into equation (27) implies\newline
\begin{eqnarray}
&-&\sum_{n=0}^{\infty }\left[ U^{(n)^{^{\prime }}}\bar{l}^{-n/2}+G^{(n)^{^{%
\prime }}}\bar{l}^{-(n+1)/2}\right]  \nonumber \\
&-&\sum_{n=0}^{\infty }\sum_{p=0}^{\infty }\left[ U^{(n)}U^{(p)}\bar{l}%
^{-(n+p)/2}+G^{(n)}G^{(p)}\bar{l}^{-(n+p+2)/2}+2U^{(n)}G^{(p)}\bar{l}%
^{-(n+p+1)/2}\right]  \nonumber \\
&+&\sum_{n=0}^{\infty }v^{(n)}\bar{l}^{-n/2}=\frac{1}{\bar{l}}\left( \beta
^{2}-\frac{1}{4}+\lambda ^{(0)}\right) +\sum_{n=2}^{\infty }\lambda ^{(n-1)}%
\bar{l}^{-n},
\end{eqnarray}
\newline
where primes of $U^{(n)}(x)$ and $G^{(n)}(x)$ denote derivatives with
respect to $x$. Equating terms of same order in $\bar{l}$ one obtains\newline
\begin{equation}
-[U^{(0)^{^{\prime }}}+U^{(0)}U^{(0)}]+v^{(0)}=0,
\end{equation}
\newline
\begin{equation}
U^{(0)^{^{\prime }}}(x)=D_{1,0}\text{ \ \ \ \ \ \ ; \ \ }D_{1,0}\text{ \ }%
=-\omega /2\text{\ \ \ \ \ \ \ }
\end{equation}

Integration over $dx$ yields 
\begin{equation}
U^{(0)}(x)=-\omega x/2\text{\ \ \ }
\end{equation}

Similarly,

\begin{equation}
-[U^{(1)^{^{\prime }}}+G^{(0)^{^{\prime
}}}]-2U^{(0)}U^{(1)}-2U^{(0)}G^{(0)}+v^{(1)}=0,
\end{equation}

\begin{equation}
U^{(1)}(x)=0,
\end{equation}

\begin{equation}
G^{(0)}(x)=C_{0,0}+C_{1,0}x^{2},
\end{equation}

\begin{equation}
C_{1,0}=-\frac{B_{1}}{w},
\end{equation}

\begin{equation}
C_{0,0}=\frac{1}{w}(2C_{1,0}-w),
\end{equation}

\begin{equation}
B_{1}=-4+\frac{\rho _{o}^{5}}{6Q}\frac{d^{3}}{d\rho _{o}^{3}}V(\rho _{o}),
\end{equation}
\newline
\begin{eqnarray}
&&-[U^{(2)^{^{\prime }}}+G^{(1)^{^{\prime
}}}]-\sum_{n=0}^{2}U^{(n)}U^{(2-n)}-G^{(0)}G^{(0)}  \nonumber \\
&&-2\sum_{n=0}^{1}U^{(n)}G^{(1-n)}+v^{(2)}=\beta ^{2}-\frac{1}{4}+\lambda
^{(0)},
\end{eqnarray}

\begin{equation}
U^{(2)}(x)=D_{1,2}x+D_{2,2}x^{3},
\end{equation}

\begin{equation}
G^{(1)}(x)=0,
\end{equation}

\begin{equation}
D_{2,2}=\frac{1}{w}(C_{1,0}^{2}-B_{2})
\end{equation}
\newline
\begin{equation}
D_{1,2}=\frac{1}{w}(3D_{2,2}+2C_{0,0}C_{1,0}-6\beta ),
\end{equation}
\newline
\begin{equation}
B_{2}=5+\frac{\rho _{o}^{6}}{24Q}\frac{d^{4}}{d\rho _{o}^{4}}V(\rho _{o}),
\end{equation}
\newline
\begin{equation}
\lambda ^{(0)}=-(D_{1,2}+C_{0,0}^{2}).
\end{equation}
\newline
\newline
... and so on. Clearly one can calculate the energy eigenvalues and the
eigenfunctions from the knowledge of C$_{n,m}$ and D$_{n,m}$ in a
hierarchical manner. However, it is for the convenience of this study to
conclude the procedure here and give the energy eigenvalues and
eigenfunctions as\newline
\begin{equation}
E=E^{(-2)}\bar{l}^{2}+E^{(0)}+E^{(1)}\bar{l}^{-1}+E^{(2)}\bar{l}^{-2}+O(\bar{%
l}^{-3}),
\end{equation}
\newline
\begin{eqnarray}
U^{^{\prime }}(x(\rho )) &=&U^{(0)}+[G^{(0)}+U^{(1)}]\bar{l}%
^{-1/2}+[G^{(1)}+U^{(2)}]\bar{l}^{-1}  \nonumber \\
&&+[G^{(2)}+U^{(3)}]\bar{l}^{-3/2}+[G^{(3)}+U^{(4)}]\bar{l}%
^{-2}+[G^{(4)}+U^{(5)}]\bar{l}^{-5/2}  \nonumber \\
&&+[G^{(5)}+U^{(6)}]\bar{l}^{-3}+O(\bar{l}^{-7/2}).
\end{eqnarray}
\newline
From which we find\newline
\begin{equation}
\Psi _{0}(x(\rho ))=exp\left( \int U^{^{\prime }}(x)dx\right) .
\end{equation}
\newline

\section{ Applications, results and discussions}

In this section the above analytical expressions of 2D - PSLET are
investigated through the 2D Coulomb, harmonic oscillator and the hybrid of
the two potentials.

\subsection{The Coulomb potential $V(\protect\rho) = - 2/\protect\rho$}

Following 2D - PSLET theory one obtains, for nodeless states,\newline
\begin{equation}
E^{(-2)}\bar{l}^2 = -\frac{1}{\bar{l}^{2}}~~;~~\bar{l} = |m|+1/2~~,~~ \rho_o=%
\bar{l}^2,
\end{equation}
\newline
\begin{equation}
E^{(0)} = E^{(1)}\bar{l}^{-1} = E^{(2)}\bar{l}^{-2} = \cdots = 0,
\end{equation}
\newline
\begin{equation}
U(x)=-\bar{l}y+\bar{l}\left(y-\frac{y^2}{2}+\frac{y^3}{3}-\frac{y^4}{4} +%
\frac{y^5}{5}-\frac{y^6}{6}+\frac{y^7}{7}-\frac{y^8}{8}\right),
\end{equation}
\newline
where $y = x \bar{l}^{-1/2}$. It is obvious that the second term in equation
(85) is the infinite geometric series expansion for $ln(1+y)^{\bar{l}}$.
Equation (85) can thus be approximated by\newline
\begin{equation}
U(x) \simeq - \bar{l} y + ln(1+y)^{\bar{l}},
\end{equation}
\newline
which in turn implies that\newline
\begin{equation}
\Psi_{0,m} (\rho) = \left(\frac{\rho}{\rho_o}\right)^{\bar{l}} e^{\bar{l}}
e^{-\bar{l}\rho/\rho_o}.
\end{equation}
\newline
Hence the nodeless radial parts of the wave functions are\newline
\begin{equation}
R_{0,m} (\rho) = N \rho^{\bar{l}-1} e^{-\bar{l} \rho/\rho_o},
\end{equation}
\newline
where $N$ is the normalization constant given by the relation\newline
\begin{equation}
N=\frac{(2\bar{l} /\rho_o)^{\bar{l}+1/2}}{\sqrt{(2\bar{l})!}}.
\end{equation}
\newline
Equation (88) evidently gives the exact expressions for the normalized
nodeless radial parts of the wave functions.

Finally, the exact energy eigenvalues for any nodeless orbital state are%
\newline
\begin{equation}
E_{0,m} = - (|m|+1/2)^{-2}.
\end{equation}
\newline

\subsection{ The harmonic oscillator potential $V(\protect\rho) = \protect%
\gamma^{2} \protect\rho^{2}/4$}

For this potential, 2D - PSLET procedure yields, for nodeless states,\newline
\begin{equation}
E^{(-2)}\bar{l}^2 = \gamma \bar{l} ~~~;~~~\bar{l} = |m|+1,
\end{equation}
\newline
\begin{equation}
E^{(0)} = E^{(1)}\bar{l}^{-1} = E^{(2)}\bar{l}^{-2} = \cdots = 0,
\end{equation}
\newline
\begin{eqnarray}
U(x)&=&-\frac{1}{2}\left(y-\frac{y^2}{2}+\frac{y^3}{3}-\frac{y^4}{4} +\frac{%
y^5}{5}-\frac{y^6}{6}\right)  \nonumber \\
&&+\bar{l}\left(y-\frac{y^2}{2}+\frac{y^3}{3}-\frac{y^4}{4}+\frac{y^5}{5} -%
\frac{y^6}{6}+\frac{y^7}{7}-\frac{y^8}{8}\right)  \nonumber \\
&&-\bar{l}\frac{y^2}{2}-\bar{l}y,
\end{eqnarray}
\newline
Obviously the terms in brackets are the infinite geometric series expansions
for $ln(1+y)$. Equation (93) can thus be approximated by\newline
\begin{equation}
U(x) \simeq ln(1+y)^{-1/2} + ln(1+y)^{\bar{l}} - \bar{l}y - \bar{l}\frac{y^2%
}{2},
\end{equation}
\newline
which in turn implies that\newline
\begin{equation}
\Psi_{0,m} (\rho) = \left(\frac{\rho}{\rho_o}\right)^{\bar{l}-1/2}
e^{-\gamma \rho^2 /4}~~~;~~~\rho_o=\sqrt{2\bar{l}/\gamma}.
\end{equation}
\newline
Hence, the nodeless radial parts of the wave functions are\newline
\begin{equation}
R_{0,m} (\rho) = N \rho^{\bar{l}-3/2} e^{-\gamma \rho^2 / 4},
\end{equation}
\newline
where the normalization constant $N$ is given through the relation\newline
\begin{equation}
N^2 = \frac{2^{\bar{l}+1/2} (\gamma/2)^{\bar{l}-1/2}}{1 \cdot 3 \cdot 5
\cdot \cdots \cdot (2\bar{l}-2)} \sqrt{\frac{\gamma}{2\pi}}.
\end{equation}
\newline
Equation (96) clearly yields the exact expressions for the normalized
nodeless radial parts of the wave functions. The exact energy eigenvalues
for any nodeless state are given by\newline
\begin{equation}
E_{0,m} = \gamma(|m|+1).
\end{equation}
\newline

\subsection{ A hybrid of Coulomb and oscillator potentials}

Perhaps the most interesting form of such a hybrid model is\newline
\begin{equation}
V(\rho )=m\gamma -\frac{2}{\rho }+\frac{\gamma ^{2}\rho ^{2}}{4},
\end{equation}
\newline
where $m$ is the magnetic quantum number. This potential (99) describes a 2D
electron gas in the x-y plane in the presence of a hydrogenic potential,
representing the interaction between a conduction electron and a donor
impurity, in a magnetic field in the z - direction. It also resembles the
potential model that describes a donor impurity in a single quantum dot
moving in a magnetic field in the z - direction [4-14].

It is obvious from equation (17), along with (18) and (19), that one can
hardly find an analytical solution for $\rho_o$, in terms of $\gamma$ and $m$%
, using the potential in (99). One has therefore to appeal to numerical
techniques to solve for $\rho_o$ for each $\gamma$ and $m$. Once $\rho_o$ is
found, the energy eigenvalues as well as eigenfunctions can be obtained
through the 2D-PSLET theory described in section 2.

In tables 1-4 we list the energy eigenvalues in such a way that the
contribution of each energy correction is made clear. The tables show that
2D-PSLET results are rapidly convergent. It is also evident that the energy
eigenvalues for states with larger $|m|$ converge more rapidly than those
with smaller $|m|$. Such a tendency was obvious from the very moment of the
invention of the deceptive perturbation parameter $1/\bar{l}$, $\bar{l}%
=|m|-\beta$ and $\beta$ is always negative.

It could be interesting to know that the computation time for all $\rho_o$'s
needed for the entries in tables 1-4 is less than 30 sec, and for each entry
is at most 20 sec, including the eigenfunctions for each of them. $\rho_o$
is computed using EUREKA and the eigenvalues as well as eigenfunctions, in
terms of $\rho$, are computed using REDUCE 3.4 on a standard Pentium PC.

In figure 1 ( Available from authour upon request ) , 2D-PSLET results
(solid curve) for the 1S-state compare excellently with the others in the
weak magnetic field regime. In the strong field regime, 2D-PSLET results
fall in between the perturbation (small dashes) [8] and direct numerical
integration (solid curve connecting solid circles) [12], on which the series
expansion results (solid circles on the solid curve) [9] are located. SLET
and SLNT [3] results are presented by the long dashed curve. 2D-PSLET, SLET
and SLNT results are, however, unique in their tendency to approach the
strong and the weak magnetic field perturbation results [8]. The
perturbation coupling constants were appropriately defined in these regimes.
Likewise, we believe, it should be the tendency of the results of any
approximation technique.

Figures 2-4 ( Available from authour upon request ) show that 2D-PSLET
results (empty squares) are in exact agreements with those of the numerical
solutions of the Schr\"{o}dinger equation [10]. The best fit line of
2D-PSLET results exactly overlaps with that of the lowest two - point
quasifractional approximation [10].

It should be mentioned that, for all states considered above, 2D-PSLET
results for the energies, excluding the third - order correction, are
exactly the same as those of SLET [3] and SLNT [3,5]. It is only because of
the third - order correction, $E^{(2)}/\bar{l}^{2}$, that the 2D-PSLET
results compared better, with the numerical results, than those of SLET and
SLNT.

\section{Conclusions and Remarks}

In this work, the psuedoperturbative shifted - $l$ expansion technique
(PSLET) was introduced to find nodeless states of 2D Schr\"{o}dinger
equation with arbitrary cylindrically symmetric potentials. Exact energy
eigenvalues and eigenfunctions for 2D Coulomb and harmonic oscillator
potentials were reproduced. Also, exact energy eigenvalues, compared to
numerical ones [10], were obtained for the hybrid of the 2D Coulomb and
oscillator potentials. The accuracy and rapid convergence of 2D-PSLET
results are satisfactory and fascinating. The analytical results, tables,
and figures clearly bear this out.

Finally, some observations concerning the attendant 2D-PSLET are in order.
It is highly accurate and rapidly convergent, thus efficient with respect to
computation time. Within the same procedure, it produces both eigenvalues
and eigenfunctions. It puts no constraints on the coupling constants of the
potential involved. It is to be understood as being an expansion through any
existing quantum number in the centrifugal - like term of any Scr\"odinger -
like equation, equation (1).

\newpage 
\begin{table}[tbp]
\caption{PSLET results for 1S-state energies (in effective Rydberg units) of
the 2D hydrogenic levels in a magnetic field , where $EN_{0}=\bar{l}%
^{2}E^{(-2)}$, $EN_{1}=EN_{0}+E^{(0)}$, $EN_{2}=EN_{1}+E^{(1)}/\bar{l}$, and 
$EN_{3}=EN_{2}+E^{(2)}/\bar{l}^{2}$.}
\begin{center}
\vspace{1cm} 
\begin{tabular}{|c|c|c|c|c|}
\hline\hline
$\gamma$ & $EN_0$ & $EN_1$ & $EN_2$ & $EN_3$ \\ \hline
0 & -4.0 & -4.0 & -4.0 & -4.0 \\ 
1 & -3.936821 & -3.907949 & -3.910053 & -3.910538 \\ 
2 & -3.738814 & -3.651631 & -3.677083 & -3.673240 \\ 
3 & -3.378562 & -3.284620 & -3.356622 & -3.314095 \\ 
4 & -2.810985 & -2.864030 & -2.934645 & -2.900042 \\ 
5 & -2.024429 & -2.380567 & -2.401652 & -2.496601 \\ 
6 & -1.141379 & -1.784237 & -1.840713 & -1.971307 \\ 
7 & -0.286914 & -1.115087 & -1.264520 & -1.373630 \\ 
8 & 0.525016 & -0.421911 & -0.660921 & -0.761824 \\ 
9 & 1.311535 & 0.279818 & -0.032861 & -0.139894 \\ 
10 & 2.085744 & 0.987287 & 0.614648 & 0.494518 \\ 
12 & 3.623940 & 2.419691 & 1.954631 & 1.802345 \\ 
20 & 9.888892 & 8.387814 & 7.701685 & 7.438987 \\ 
28 & 16.38730 & 14.65978 & 13.83081 & 13.48993 \\ 
36 & 23.06128 & 21.13973 & 20.19640 & 19.79357 \\ 
40 & 26.44867 & 24.43836 & 23.44415 & 23.01409 \\ \hline\hline
\end{tabular}
\end{center}
\end{table}
\newpage

\begin{table}[tbp]
\caption{PSLET results for 1S-state energies (in effective Rydberg units) of
the 2D hydrogenic levels in a magnetic field , where $EN_0 = \bar{l}^2
E^{(-2)}$, $EN_1 = EN_0 + E^{(0)}$, $EN_2 = EN_1 + E^{(1)}/\bar{l}$, $EN_3 =
EN_2 + E^{(2)}/\bar{l}^2$, and ${\protect\gamma}^{^{\prime}}$ $= \protect%
\gamma/(1+\protect\gamma)$.}
\begin{center}
\vspace{1cm} 
\begin{tabular}{|c|c|c|c|c|}
\hline\hline
${\gamma}^{^{\prime}}$ & $EN_0$ & $EN_1$ & $EN_2$ & $EN_3$ \\ \hline
0.0 & -4.0 & -4.0 & -4.0 & -4.0 \\ 
0.1 & -3.999228 & -3.998843 & -3.998843 & -3.998830 \\ 
0.2 & -3.996091 & -3.994148 & -3.994156 & -3.994162 \\ 
0.3 & -3.988498 & -3.982838 & -3.982914 & -3.982940 \\ 
0.4 & -3.972089 & -3.958670 & -3.959106 & -3.959233 \\ 
0.5 & -3.936821 & -3.907949 & -3.910053 & -3.910542 \\ 
0.6 & -3.855896 & -3.797597 & -3.807128 & -3.807128 \\ 
0.7 & -3.638726 & -3.538277 & -3.578849 & -3.566445 \\ 
0.8 & -2.810985 & -2.864030 & -2.934645 & -2.900042 \\ \hline\hline
\end{tabular}
\end{center}
\end{table}
\newpage

\begin{table}[tbp]
\caption{PSLET results for 2P$^-$ - state energies (in effective Rydberg
units) of the 2D hydrogenic levels in a magnetic field , where $EN_0 = \bar{l%
}^2 E^{(-2)}$, $EN_1 = EN_0 + E^{(0)}$, $EN_2 = EN_1 + E^{(1)}/\bar{l}$, $%
EN_3 = EN_2 + E^{(2)}/\bar{l}^2$, and ${\protect\gamma}^{^{\prime}}$ $= 
\protect\gamma/(1+\protect\gamma)$.}
\begin{center}
\vspace{1cm} 
\begin{tabular}{|c|c|c|c|c|}
\hline\hline
${\gamma}^{^{\prime}}$ & $EN_0$ & $EN_1$ & $EN_2$ & $EN_3$ \\ \hline
0.0 & -4/9 & -4/9 & -4/9 & -4/9 \\ 
0.1 & -0.525152 & -0.523702 & -0.523991 & -0.523943 \\ 
0.2 & -0.556955 & -0.562580 & -0.562780 & -0.562971 \\ 
0.3 & -0.542085 & -0.561985 & -0.563083 & -0.563600 \\ 
0.4 & -0.481468 & -0.514474 & -0.518498 & -0.519116 \\ 
0.5 & -0.355319 & -0.400522 & -0.408080 & -0.409164 \\ 
0.6 & -0.113944 & -0.172226 & -0.183710 & -0.185583 \\ 
0.7 & 0.366646 & 0.292098 & 0.275860 & 0.272867 \\ 
0.8 & 1.478312 & 1.379563 & 1.356515 & 1.351825 \\ \hline\hline
\end{tabular}
\end{center}
\end{table}
\newpage

\begin{table}[tbp]
\caption{PSLET results for 3D$^-$ - state energies (in effective Rydberg
units) of the 2D hydrogenic levels in a magnetic field , where $EN_0 = \bar{l%
}^2 E^{(-2)}$, $EN_1 = EN_0 + E^{(0)}$, $EN_2 = EN_1 + E^{(1)}/\bar{l}$, $%
EN_3 = EN_2 + E^{(2)}/\bar{l}^2$, and ${\protect\gamma}^{^{\prime}}$ $= 
\protect\gamma/(1+\protect\gamma)$.}
\begin{center}
\vspace{1cm} 
\begin{tabular}{|c|c|c|c|c|}
\hline\hline
${\gamma}^{^{\prime}}$ & $EN_0$ & $EN_1$ & $EN_2$ & $EN_3$ \\ \hline
0.0 & -4/25 & -4/25 & -4/25 & -4/25 \\ 
0.1 & -0.257041 & -0.260346 & -0.260476 & -0.260508 \\ 
0.2 & -0.264976 & -0.272340 & -0.273106 & -0.273176 \\ 
0.3 & -0.225665 & -0.236222 & -0.237588 & -0.237735 \\ 
0.4 & -0.134531 & -0.148204 & -0.150156 & -0.150392 \\ 
0.5 & 0.031455 & 0.014366 & 0.011785 & 0.011447 \\ 
0.6 & 0.325970 & 0.304776 & 0.301450 & 0.300990 \\ 
0.7 & 0.881999 & 0.855342 & 0.851037 & 0.850414 \\ 
0.8 & 2.115430 & 2.080314 & 2.074507 & 2.073635 \\ \hline\hline
\end{tabular}
\end{center}
\end{table}
\newpage

\end{document}